\documentclass{article}
\usepackage[T1]{fontenc} 
\usepackage[utf8]{inputenc} 
\usepackage{ismir,amsmath,cite,url}
\usepackage{graphicx}
\usepackage{color}
\usepackage{url}

\title{MIDI Passage Retrieval Using Cell Phone Pictures of Sheet Music}

\multauthor
{Daniel Yang${^*}$ \hspace{.5cm} Thitaree Tanprasert${^*}$ \thanks{$^*$The first two authors had equal contribution.} \hspace{.5cm} Teerapat Jenrungrot \hspace{.5cm} Mengyi Shan \hspace{.5cm} TJ Tsai} {
  Harvey Mudd College, Claremont, CA USA \\
{\tt\small \{dhyang, ttanprasert, mjenrungrot, mshan, ttsai\}@hmc.edu}
}
\def\authorname{Daniel Yang, Thitaree Tanprasert, Teerapat Jenrungrot, Mengyi Shan, TJ Tsai}

\begin{document}

\maketitle
\begin{abstract}
This paper investigates a cross-modal retrieval problem in which a user would like to retrieve a passage of music from a MIDI file by taking a cell phone picture of a physical page of sheet music.  While audio--sheet music retrieval has been explored by a number of works, this scenario is novel in that the query is a cell phone picture rather than a digital scan.  To solve this problem, we introduce a mid-level feature representation called a bootleg score which explicitly encodes the rules of Western musical notation.  We convert both the MIDI and the sheet music into bootleg scores using deterministic rules of music and classical computer vision techniques for detecting simple geometric shapes.  Once the MIDI and cell phone image have been converted into bootleg scores, we estimate the alignment using dynamic programming.  The most notable characteristic of our system is that it does test-time adaptation and has no trainable weights at all---only a set of about 30 hyperparameters.  On a dataset containing 1000 cell phone pictures taken of 100 scores of classical piano music, our system achieves an F measure score of $.869$ and outperforms baseline systems based on commercial optical music recognition software.
\end{abstract}
\section{Introduction}
\label{sec:intro}

Consider this scenario:  A person is practicing at the piano, and would like to know what a particular passage of music sounds like.  She takes a cell phone picture of a portion of the physical sheet music in front of her, and is immediately able to hear what those lines of music sound like.  

In this paper, we explore the feasibility of such an application where we assume that the piece is known and a MIDI file of the piece is available.  Our goal is to retrieve a passage of music from a MIDI file using a cell phone image as a query.  This is a cross-modal retrieval scenario.

Several works have investigated the correspondence between audio and sheet music images.  There are two general approaches to the problem.  The first approach is to use an existing optical music recognition (OMR) system to convert the sheet music into a symbolic (MIDI-like) representation, to compute chroma-like features, and then to compare the resulting sequences to chroma features extracted from the audio.  This approach has been applied to synchronizing audio and sheet music \cite{damm2012digital, DammFKMC08_MultimodalPresentationofMusic_ICMI, KurthMFCC07_AutomatedSynchronization_ISMIR, ThomasFMC12_LinkingSheetMusicAudio_DagstuhlFU, izmirli2012bridging}, identifying audio recordings that correspond to a given sheet music representation \cite{FremereyMKC08_AutomaticMapping_ISMIR}, and finding the audio segment corresponding to a fragment of sheet music \cite{FremereyCME09_SheetMusicID_ISMIR}.  A different approach has been explored in recent years: convolutional neural networks (CNNs).  This approach attempts to learn a multimodal CNN that can embed a short segment of sheet music and a short segment of audio into the same feature space, where similarity can be computed directly.  This approach has been explored in the context of online sheet music score following \cite{dorfer2016live}, sheet music retrieval given an audio query \cite{dorfer2016towards, dorfer2017learning, dorfer2018end, dorfer2018tismir}, and offline alignment of sheet music and audio \cite{dorfer2017learning}.  Dorfer et al. \cite{dorfer2018learning} have also recently shown promising results formulating the score following problem as a reinforcement learning game.  See \cite{mueller2019cross} for a recent overview of work in this area.

The key novelty in our scenario is the fact that the queries are cell phone images.  All of the works described above assume that the sheet music is either a synthetically rendered image or a digital scan of printed sheet music.  In recent years, a few works have begun to explore optical music recognition (OMR) on camera-based musical scores \cite{calvo2018camera, bui2014staff, vo2016mrf, blanes2017camera, vo2014distorted}.  Here, we explore the use of cell phone images of sheet music for \emph{retrieval}.  Cell phone images provide a natural and convenient way to retrieve music-related information, and this motivates our current study.

The main conceptual contribution of this paper is to introduce a mid-level feature representation called a bootleg score which explicitly encodes the conventions of Western musical notation.  As we will show, it is possible to convert MIDI into a bootleg score using the rules of musical notation, and to convert the cell phone image into a bootleg score using classical computer vision techniques for detecting simple geometrical shapes.  Once we have converted the MIDI and cell phone image into bootleg feature space, we can estimate the alignment using subsequence DTW.  The most notable characteristic of our system is that it does test-time adaptation and \emph{contains no trainable weights at all}---only a set of approximately $30$ hyperparameters.  In the remainder of this paper, we will describe the system and present our experimental results.


\section{System Description}
\label{sec:systemDescr}

\begin{figure}
	\includegraphics[width=\columnwidth]{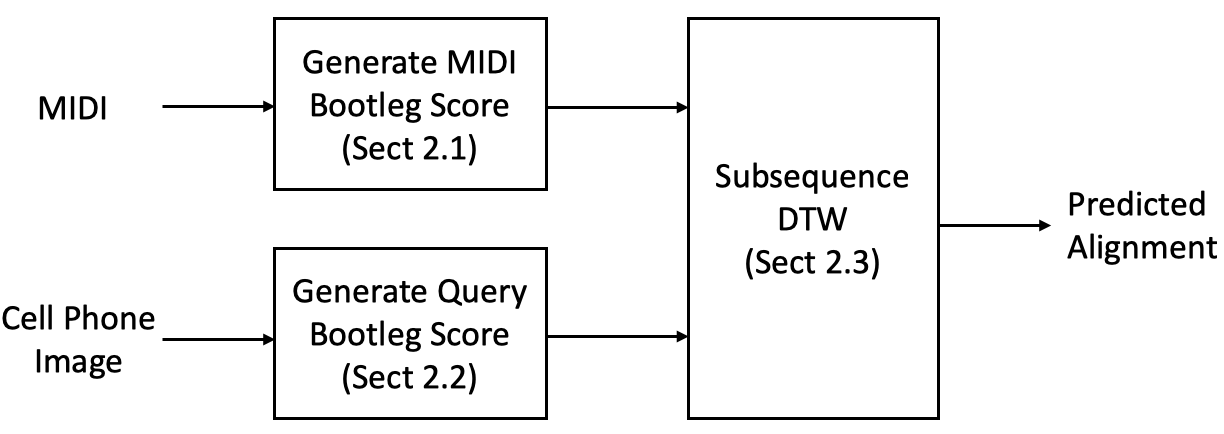}
	\caption{Block diagram of the proposed system.}
	\label{fig:systemOverview}
\end{figure}

Our system takes two inputs: a cell phone picture of a page of sheet music and a MIDI file of the corresponding piece.  The output of the system is a prediction of the time segment in the MIDI file that matches the lines of sheet music shown in the cell phone picture.  Note that in this problem formulation, we assume that the piece is known, and that we are trying to identify the matching passage of music in the piece.  In our study, we focus on piano music.

Our approach has three main components, which are shown in Figure \ref{fig:systemOverview}.  The first two components convert the MIDI and cell phone image into a representation which we call a bootleg score.  A bootleg score is a very low-dimensional representation of music which is a hybrid between sheet music and MIDI.  It is a manually designed feature space that explicitly encodes the rules of Western musical notation.  The third component is to temporally align the two bootleg scores using subsequence DTW.  These three components will be discussed in the next three subsections.\footnote{Our code is available at \url{https://github.com/tjtsai/SheetMidiRetrieval}}

\subsection{Generating MIDI Bootleg Score}

\begin{figure}
	\includegraphics[width=\columnwidth]{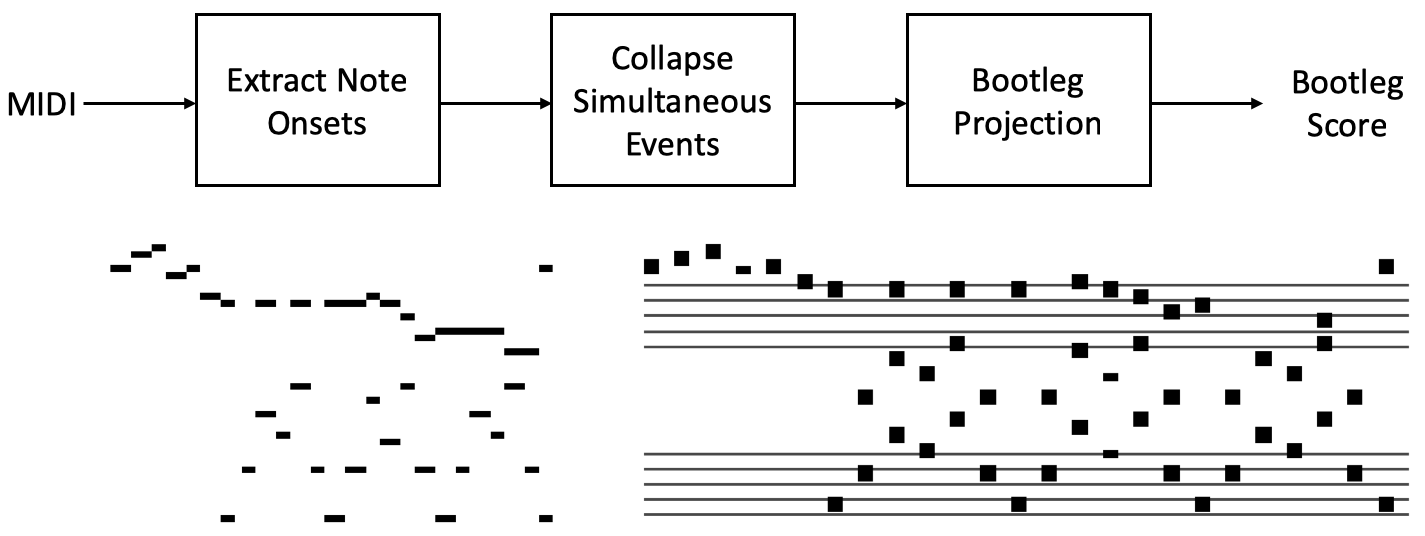}
	\caption{Overview of generating the MIDI bootleg score (Section 2.1).  Below the block diagram, a short MIDI passage (left) and its corresponding bootleg score are shown.}
	\label{fig:generateMidiBootleg}
\end{figure}

Generating the MIDI bootleg score consists of the three steps shown in Figure \ref{fig:generateMidiBootleg}.  The first step is to extract a list of all individual note onsets.  The second step is to cluster the note onsets into groups of simultaneous note events.  After this second step, we have a list of note events, where each note event consists of one or more simultaneous note onsets.  The third step is to project this list of note events into the bootleg feature space.  

The bootleg feature representation can be thought of as a very crude version of sheet music (thus the name ``bootleg score").  It asks the question, ``If I were to look at the sheet music corresponding to this MIDI file, where would the notehead for each note onset appear among the staff lines?"  Note that there is ambiguity when mapping from a MIDI note value to a position in a staff line system.  For example, a note onset with note value 60 (C4) could appear in the sheet music as a C natural or a B sharp,\footnote{It could also appear as a D double flat, but we do not consider double sharps or double flats since they occur relatively infrequently.}  and it could also appear in the right hand staff (i.e. one ledger line below a staff with treble clef) or the left hand staff (i.e. one ledger line above a staff with bass clef).  The bootleg feature representation handles ambiguity by simply placing a notehead at all possible locations.  The bootleg score is a binary image containing only these floating noteheads.

The bootleg score is a very low dimensional representation.  Along the vertical dimension, it represents each staff line location as a single bootleg pixel (which we will refer to as a ``bixel'' to differentiate between high-dimensional raw image pixels and low-dimensional bootleg score pixels).  For example, two adjacent staff lines would span three bixels: two bixels for the staff lines and one bixel for the position in between.  The bootleg score contains both right hand and left hand staves, similar to printed piano sheet music.  In total, the bootleg score is 62 bixels tall.  Along the horizontal dimension, we represent each simultaneous note event as a single bixel column.  We found through experimentation that a simple modification improves performance in the alignment stage (Section \ref{subsec:alignment}): we simply repeat each bixel column twice and insert an empty bixel column between each simultaneous note event.  This gives the system more flexibility to deal with noisy bixel columns in the alignment stage.

The resulting MIDI bootleg score is a $62 \times 3N$ binary matrix, where $N$ is the number of simultaneous note events in the MIDI file.\footnote{The factor of $3$ comes from the filler and repetitions.} Figure \ref{fig:generateMidiBootleg} shows an example bootleg score.  The staff lines are included as a visualization aid, but are not present in the bootleg feature representation.

\subsection{Generating Query Bootleg Score}
\label{sec:genQueryBootleg}

The second main component of our system (Figure \ref{fig:systemOverview}) is to convert the cell phone image into a bootleg score representation.  Unlike the MIDI representation, the image does not explicitly encode any information about the notes, so we will have to estimate this information from the raw image.



Our general approach rests on two key insights.  The first key insight is that we can identify where we are in the piece if we can detect just three things: filled noteheads, staff lines, and bar lines.  Because these objects are simple geometrical shapes, classical computer vision tools are sufficient to detect them (Section \ref{subsec:noteheadDetect}--\ref{subsec:barlineDetect}).  The second key insight is that we know a priori that these three objects will occur many times in the image.  This opens up the possibility of test-time adaptation, where we can use a very simple notehead detector to identify some of the noteheads in the image, and then use those detected instances to learn a more accurate notehead template at test time.  This is generally not possible with large object detection and classification scenarios like the ImageNet competition \cite{deng2009imagenet, russakovsky2015imagenet}.

\begin{figure}
\includegraphics[width=\columnwidth]{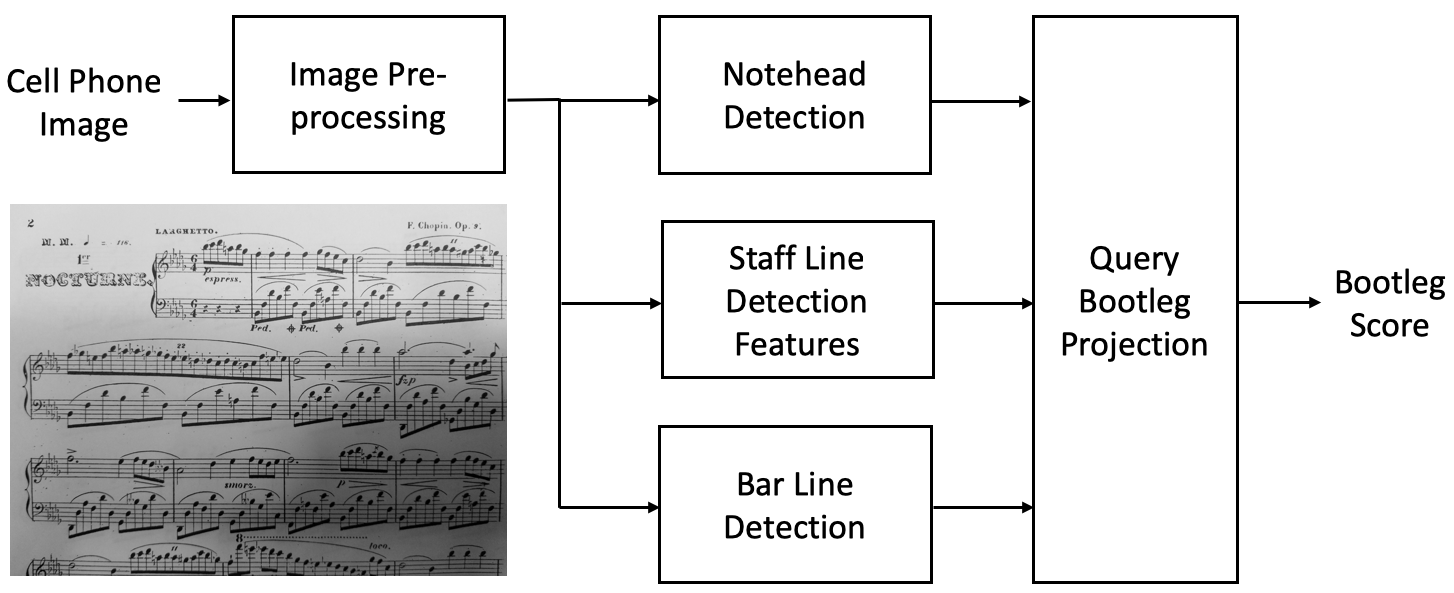}
\caption{Overview of generating the query bootleg score (Sections 2.2.1--2.2.5).  The cell phone image shown will serve as a running example throughout the paper.}
\label{fig:generateQueryBootleg}
\end{figure}

Our method for generating the cell phone image bootleg score has five parts, which are shown in Figure \ref{fig:generateQueryBootleg}.  These will be described in the next five subsections.

\subsubsection{Image Pre-processing}

The preprocessing consists of three operations: (1) converting the image to grayscale, (2) resizing the image to a maximum dimension of $1000$ pixels while retaining the same aspect ratio, and (3) removing background lighting by blurring the image and then subtracting the blurred image from the non-blurred image.

\subsubsection{Notehead Detection}
\label{subsec:noteheadDetect}

\begin{figure}
	\includegraphics[width=\columnwidth]{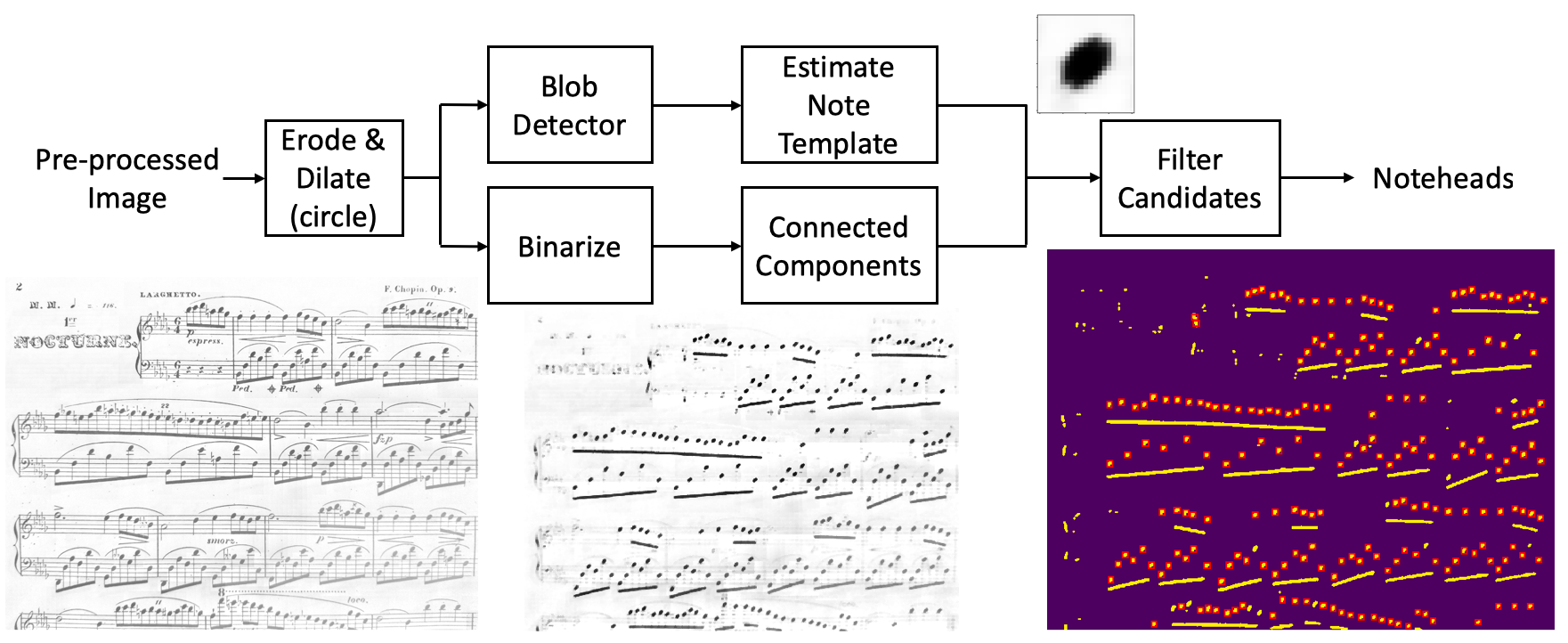}
	\caption{Overview of notehead detection (Section \ref{subsec:noteheadDetect}).  The images at bottom show the pre-processed image before (left) and after (center) erosion \& dilation, and the detected noteheads (right).}
	\label{fig:noteheadDetection}
\end{figure}

The goal of the notehead detection stage in Figure \ref{fig:generateQueryBootleg} is to predict a bounding box around every filled notehead in the cell phone image.  Note that we do not attempt to detect non-filled noteheads (i.e. half-notes, dotted half notes, whole notes).  The basic premise of our approach is that filled noteheads are much easier to detect, and they also generally occur much more frequently than half or whole notes.  The notehead detection consists of the steps shown in Figure \ref{fig:noteheadDetection}.  We will explain these steps in the next four paragraphs.

The first step is to perform erosion and dilation of the pre-processed image with a circular morphological filter.  The erosion replaces each pixel with the whitest pixel in a circular region centered around the pixel.  This operation removes any objects that consist of thin lines, and it only passes through contiguous dense regions of black pixels.  The dilation takes the resulting image and replaces each pixel with the blackest pixel in a circular region center around the pixel.  This operation restores any objects that survived the erosion back to their original size.  Figure \ref{fig:noteheadDetection} shows an example of an image after erosion and dilation (center image).

Next, we describe the processing in the upper path of Figure \ref{fig:noteheadDetection}.  We take the eroded and dilated image and apply simple blob detection.  We use the simple blob detector in OpenCV with default parameter settings, except that we specify a minimum and maximum area in order to specify the rough size of object we expect.  We then take crops of the (eroded and dilated) image around the detected keypoints, and we compute the average of the cropped regions.  This average gives us an estimate of what a filled notehead looks like in this image.  Figure \ref{fig:noteheadDetection} shows an example of an estimated template (upper right).

Now we describe the processing in the lower path of Figure \ref{fig:noteheadDetection}.  We take the eroded and dilated image and binarize it using Otsu binarization \cite{otsu1979threshold}.  We then extract a list of connected component regions from the binary image, which gives us a list of candidate regions, some of which are noteheads.

The last step in notehead detection is to filter the list of candidates using our estimated notehead template.  We filter the list of candidates to only contain those regions whose height, width, height-width ratio, and area all roughly match the notehead template (within some tolerance).  We also filter the list of candidates to identify chord blocks, which often appear as a single connected component region.  When a chord block is identified, we estimate the number of notes in the chord based on its area relative to the notehead template and then perform k-means clustering to estimate individual notehead locations.

At the end of these steps, we have a list of bounding boxes around the detected notes in the cell phone image.  Figure \ref{fig:noteheadDetection} (bottom right) shows an example of the predicted notehead locations in an image.

\subsubsection{Staff Line Detection Features}
\label{subsec:stafflineDetect}

The goal of the staff line detection features stage in Figure \ref{fig:generateQueryBootleg} is to compute a tensor of features that can be used to predict staff line locations in the bootleg projection stage (Section \ref{subsec:queryBootlegProjection}).  In a cell phone picture, staff lines may not be straight lines or have equal spacing throughout the image due to the camera angle or camera lens distortions.  For these reasons, we estimate staff line locations locally rather than globally.  In other words, for every detected notehead, we make a local estimate of the staff line location and spacing in its vicinity.

\begin{figure}
	\centerline{\includegraphics[width=\columnwidth]{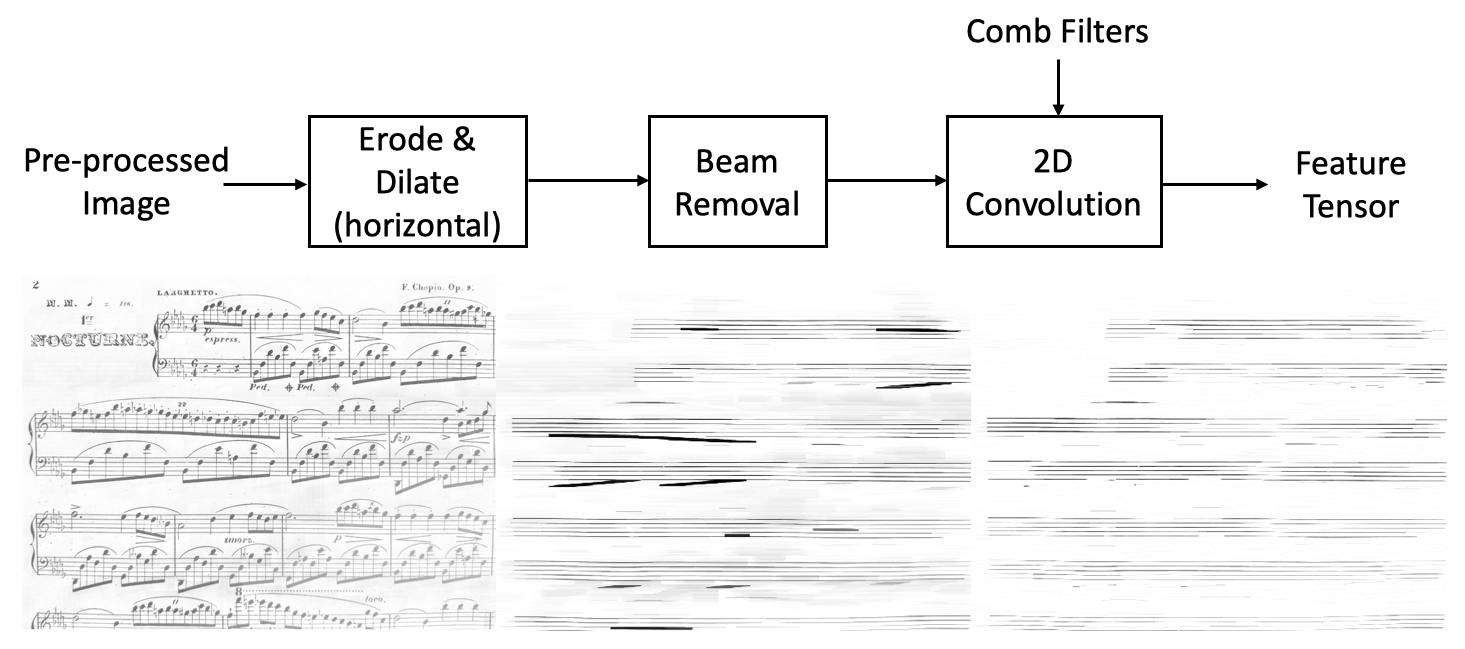}}
	\caption{Overview of staff line features computation (Section \ref{subsec:stafflineDetect}).  The images at bottom show the pre-processed image before (left) and after (middle) erosion \& dilation, and the result after removing note beams (right).  The actual feature tensor is not shown.}
	\label{fig:staffDetectBlock}
\end{figure}

The staff line detection features are computed in three steps as shown in Figure \ref{fig:staffDetectBlock}.  The first step is to perform erosion and dilation on the image with a short ($1$ pixel tall), fat morphological filter.  This removes everything except for horizontal lines.  In practice, we find that there are two types of objects that survive this operation: staff lines and horizontal note beams (e.g. the beam connecting a sequence of sixteenth notes).  The second step is to remove the note beams, as they can throw off the staff line location estimates.  Because the note beams are much thicker than staff lines, we can isolate the note beams based on their thickness and subtract them away from the image.  The third step is to convolve the resulting image with a set of comb filters.  We construct a set of tall, skinny ($1$ pixel wide) comb filters, where each comb filter corresponds to a particular staff line spacing.  The set of comb filters is selected to span a range of possible staff line sizes.  We then convolve the image (after beam removal) with each of the comb filters and stack the filtered images into a tensor.  This feature tensor $\boldsymbol{T}_{global}$ has dimension $H_{image} \times W_{image} \times N_{comb}$, where $H_{image}$ and $W_{image}$ specify the dimensions of the image and $N_{comb}$ is the number of comb filters.  Note that the third dimension corresponds to different staff line spacings.

\subsubsection{Bar Line Detection}
\label{subsec:barlineDetect}

The goal of the bar line detection stage (Figure \ref{fig:generateQueryBootleg}) is to predict a bounding box around the barlines in the cell phone image.  The bar lines are needed to correctly cluster staff lines into grand staff systems, where each grand staff consists of a right hand staff and a left hand staff.

\begin{figure}
	\centerline{\includegraphics[width=\columnwidth]{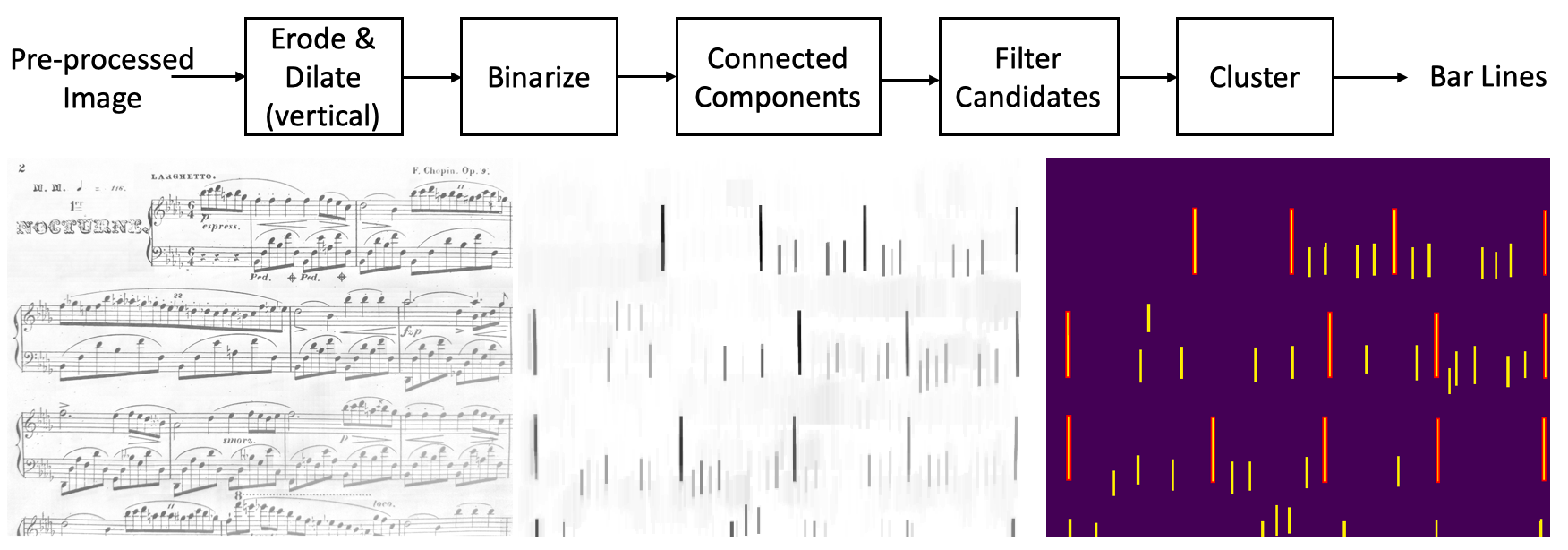}}
	\caption{Overview of bar line detection (Section \ref{subsec:barlineDetect}).  The images at bottom show the pre-processed image before (left) and after (center) erosion \& dilation, and the detected bar lines (right).}
	\label{fig:barDetectBlock}
\end{figure}

The bar line detection consists of the five steps shown in Figure \ref{fig:barDetectBlock}.  The first step is to perform erosion and dilation of the image with a tall, skinny morphological filter.  This filters out everything except vertical lines.  In practice, we find that there are three types of objects that survive this operation: bar lines, note stems, and background pixels (e.g. music stand at edges of image).  The second step is to binarize the eroded and dilated image using Otsu binarization.  The third step is to extract a list of connected component regions from the binary image.  The fourth step is to filter this list of candidates to identify bar lines.  This can be done by first filtering out regions that are too wide (e.g. background pixel regions), and then distinguishing between note stems and bar lines by finding the threshold on height that minimizes intra-class variance (which is equivalent to Otsu binarization but applied to the heights).  The fifth step is to cluster the detected bar lines into lines of music.  We do this by simply clustering any bar lines that have any vertical overlap.  Figure \ref{fig:barDetectBlock} shows this process for an example image.

At the end of the bar line detection stage, we have a prediction of the number of lines of music in the cell phone image, along with the vertical pixel range associated with each line.  Figure \ref{fig:barDetectBlock} shows an example of an image at the various stages of processing in the bar line detection.

\subsubsection{Query Bootleg Projection}
\label{subsec:queryBootlegProjection}

\begin{figure}
	\centerline{\includegraphics[width=\columnwidth]{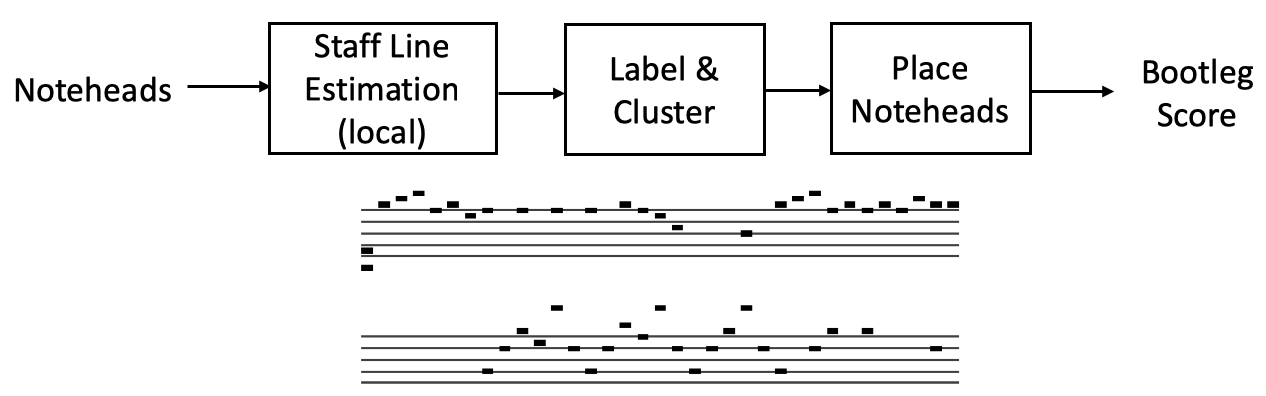}}
	\caption{Overview of query bootleg projection (Section \ref{subsec:queryBootlegProjection}).  The image at bottom shows part of the generated bootleg score for the cell phone image in Figure \ref{fig:generateQueryBootleg}.}
	\label{fig:queryBootlegProjection}
\end{figure}

The last step in Figure \ref{fig:generateQueryBootleg} is to combine the notehead, staff line, and bar line information in order to synthesize a bootleg score for the cell phone image.  This bootleg score synthesis consists of the three steps shown in Figure \ref{fig:queryBootlegProjection}.

\begin{figure}
	\centerline{\includegraphics[width=\columnwidth]{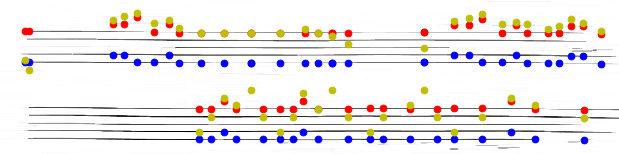}}
	\caption{A visualization of local staff line estimation.  Each yellow dot corresponds to a detected notehead, and the red and blue dots correspond to the predicted top and bottom staff lines.}
	\label{fig:stafflineEstimate}
\end{figure}

The first step is to locally estimate the staff line location and spacing for each notehead.  We do this by selecting a subset $\boldsymbol{T}_{local}$ of the staff line feature tensor $\boldsymbol{T}_{global}$ which only contains a rectangular context region around the notehead's $(x,y)$ location in the image.   This gives us a three-dimensional feature tensor $\boldsymbol{T}_{local}$ with dimension $H_{context} \times W_{context} \times N_{comb}$, where $H_{context}$ and $W_{context}$ specify the size of the context region and $N_{comb}$ specifies the number of comb filters.  We calculate the sum of features across the rows of $\boldsymbol{T}_{local}$, and then identify the maximum element in the resulting $H_{context} \times N_{comb}$ matrix.  The location of the maximum element specifies the vertical offset of the staff lines, along with the staff line size.  Figure \ref{fig:stafflineEstimate} shows a visualization of the estimated local staff line predictions for a line of music.  Yellow dots correspond to estimated notehead locations, and the red and blue dots are predictions of the top and bottom staff lines.

The second step is to label and cluster detected noteheads.  We estimate each notehead's discrete staff line location by applying simple linear regression on its local staff line coordinate system followed by quantization.  This is necessary to determine where the notehead should be placed in the bootleg score.  We also need to associate each notehead with an upper or lower staff in a specific line of music.  To do this, we first check to see if the predicted staff line system is within the vertical range of a valid line of music (see Section \ref{subsec:barlineDetect}) and, if so, if it falls in the upper or lower half of the region.  If the predicted staff line system does not fall within a valid line of music, the notehead is ignored and will not appear in the bootleg score.  The latter tends to happen with noteheads at the very top or bottom of the image, where a portion of a staff shows up but is cut off by the image boundaries.

The third step is to actually place the noteheads into the bootleg score.  We collapse the noteheads within each valid bar line region into a sequence of simultaneous note events, and then construct the bootleg score as a sequence of simultaneous note events.  Similar to the MIDI bootleg score, we repeat each simultaneous note event twice and insert a filler column between each simultaneous note event.  Figure \ref{fig:queryBootlegProjection} shows part of the bootleg score generated from the cell phone image in Figure \ref{fig:generateQueryBootleg}.

\subsection{Subsequence DTW}
\label{subsec:alignment}

The third main component of our system (Figure \ref{fig:systemOverview}) is to align the two bootleg scores using subsequence dynamic time warping (DTW).  DTW is a well-established dynamic programming technique for determining the alignment between two feature sequences.  Subsequence DTW is a variant of DTW that finds the optimal match between a shorter query segment and a subsequence of a (longer) reference segment.  For details on DTW and its variants, the reader is referred to \cite{muller2015fundamentals}.  Our cost metric computes the negative inner product between two bixel columns and then normalizes the result by the maximum of (a) the number of simultaneous noteheads in the sheet music and (b) the number of simultaneous note onsets in the MIDI.  The inner product counts how many overlapping black bixels there are between the two columns, and the normalization factor ensures that the actual cost is not biased by the number of simultaneous notes.  At the end of this stage, we have a prediction of the segment in the MIDI file that best matches the lines of sheet music shown in the cell phone image.  This is the final output of our proposed system.

\section{Experimental Setup}

The experimental setup will be described in three parts: the data, the annotations, and the evaluation metric.

The data was collected in the following manner.  We first download $100$ piano scores in PDF format from IMSLP.\footnote{\url{https://imslp.org}}  These piano scores come from $25$ well-known composers and span a range of eras and genres within the classical piano literature.  To simplify the evaluation, we select scores that do not have any repeats or structural jumps.   For each score, we then find a corresponding MIDI file from various online websites.  This gives us a total of $100$ MIDI-PDF matching pairs.  Next, we printed out the PDF scores onto physical paper, placed the sheet music pages in various locations, and took $10$ cell phone pictures of each score, spaced throughout the length of the piece.  The pictures were taken in various ambient lighting conditions (some of which triggered the flash and some of which didn't), various perspectives, and varying levels of zoom.  The pictures capture between $1$ and $4$ lines of music on a page.  We collected the data with two cell phones (iPhone 8, Galaxy S10), and all pictures were taken in landscape orientation.  As much as possible, we tried to emulate typical conditions of the application scenario.  In total, the data contains $100$ MIDI files, $100$ scores, and $1000$ cell phone images.

The data was manually annotated at the measure level.  For the MIDI files, we used \url{pretty_midi} to programatically estimate the timestamps of the downbeats in each measure, which were then manually verified and corrected.  For the cell phone images, we annotated which measures in the score were captured.  Since the images would often capture a fragment of a line of music (at the top or bottom), we adopted the convention of only annotating measures on lines of music that are fully captured in the image.  For each image, we can use these annotations to determine the matching time segment in the MIDI file.

The metric we use to evaluate our system performance is precision, recall, and F measure.  Precision is the temporal duration of overlap between the hypotheses and ground truth segments divided by the total duration of hypothesis segments.  Recall is the amount of overlap divided by the total duration of ground truth segments.  F measure is then computed as the harmonic mean of precision and recall.  In a few situations, the query perfectly matches two different sections in the score.  In these situations, we consider any perfectly matching sections of the score to be correct.

\section{Results}

We evaluate our system in the following manner.  We first randomly select $10$ out of the $100$ scores and set apart their corresponding $10 \times 10 = 100$ cell phone images as the training set.  The remaining $900$ cell phone images are set apart for testing.  Note that this train-test split has an unusually large emphasis on the test data.  The reason that we do this is because our system has no trainable weights---only hyperparameters---so the training data is really only used to determine the hyperparameter settings.  After doing iterative design on the training data and determining reasonable hyperparameter settings, we froze the system and evaluated it on the $900$ test images coming from the $90$ unseen music scores.  

We compare our system to three baselines.  The first two baseline systems are Photoscore and SharpEye, which are both commercially available OMR software.  We use the software to convert the cell phone image to a (predicted) MIDI representation and then perform subsequence DTW with chroma features.  Note that Photoscore and SharpEye were not designed to handle cell phone images, so they would sometimes fail to process the image (i.e. would throw an error).  In these situations, we simply mapped errors to a predicted time interval with $0$ duration.  The third baseline is random (informed) guessing.  We calculate the average number ($N$) of sheet music measures showing in the training images.  At test time we randomly select a time interval in the reference MIDI file spanning $N$ measures.  

\begin{table}
	\begin{center}
		\begin{tabular}{| l | l | c | c | c |}
			\hline
			System & Data & P & R & F \\
			\hline
			Random  & Test & $.152$ & $.189$ & $.169$ \\
			SharpEye & Test & $.413$ & $.091$ & $.150$ \\
			Photoscore & Test & $.692$ & $.681$ & $.687$ \\
			Bootleg & Test & $.900$ & $.840$ & $.869$ \\
			\hline
			Bootleg & Train & $.872$ & $.869$ & $.871$ \\
			\hline
		\end{tabular}
	\end{center}
	\caption{Experimental results.  The three rightmost columns show precision (P), recall (R), and F measure (F).}
	\label{tab:results}
\end{table}

Table \ref{tab:results} shows the performance of our system and the three baseline systems.  There are three things to notice about these results.  First, the baseline systems all perform poorly.  This is not a surprise, since Photoscore and SharpEye were designed to handle sheet music scans, not cell phone images.  We would expect that other OMR-based approaches that are trained on scanned sheet music would likewise perform poorly on cell phone images.  Second, the bootleg approach far outperforms the baselines.  The proposed system achieves an F measure score of $.869$ on the test set, which is far better than the highest F measure score ($.687$) among the baseline systems.  Third, the proposed system generalizes very well from the training data to the testing data.  After iterating and optimizing the system on the training data, the F measure score only fell from $.871$ (on the training data) to $.869$ (on the test data).  The reason that our system generalizes so well with such a small training data set is that our system has no trainable weights and only about $30$ hyperparameters.  Even then, many of these hyper parameters are dictated by conventions of Western musical notation for piano music.  With such a small number of parameters, we don't expect the system to suffer severely from overfitting, and indeed this is what we observe in our experiments.

\section{Analysis}

In this section we gain deeper insight into our system through two different analyses.

The first analysis is to manually investigate all of the test queries that were failures.  Here, we define a failure as having no overlap at all between the predicted time interval and the ground truth time interval.  These are instances where the system simply failed to find a reasonable match.  There were two common causes of failure.  The biggest cause of failure came from notehead detection mistakes.  The notehead detector will obviously fail on half notes and whole notes, since we only try to detect filled noteheads.  When the sheet music contains a high fraction of these notes, the system will perform poorly.  Also, the system often failed to detect chord blocks where multiple noteheads were located in close proximity to one another.  This problem is primarily due to poor hyperparameter settings, and could be mitigated by optimizing the hyperparameters over a larger, more diverse training data set.  The second cause of failure were symbols that cause the noteheads to appear in a different place than expected.  These include clef changes, octave markings, and trills.  Clef changes and octave markings could be incorporated into the MIDI bootleg score by considering all possible clef and octave changes in both right and left hand staves, but there is no immediately obvious way to address the problem of trills.  

The second analysis is to characterize run time.  Because our application is an online search, the run time is an important consideration.  Accordingly, we profiled our system to determine how long it takes to process each query, and to identify the parts of the system that are bottlenecks to improve runtime.  Note that our entire system is implemented in python with OpenCV and a custom cython-accelerated subsequence DTW function.  When each query is processed by a single $2.1$ GHz Intel Xeon processor, the average runtime is $7.6$ seconds.  When we analyze the breakdown of runtime across the major components of the system, we find that the major bottleneck is the staff line detection features stage ($92\%$ of total runtime), which primarily consists of 2-D convolutions with the set of comb filters.  This suggests one way to improve runtime: rather than using a large set of comb filters to handle a wide range of possible staff line spacings, we could explicitly estimate the staff line size and consider a much smaller set of comb filters.  If we could reduce the set of comb filters by a factor of $10$, the average time per query would be $1.3$ seconds.

\section{Conclusion}

We explore an application in which a user would like to retrieve a passage of music from a MIDI file by taking a cell phone picture of a physical page of printed sheet music.  We develop a proof-of-concept prototype and evaluate its performance on a dataset containing $1000$ cell phone pictures of $100$ different scores of classical piano music.  Our system projects both the MIDI file and the cell phone images into a low-dimensional feature representation called a bootleg score, which explicitly encodes the rules of Western musical notation.  We then align the two bootleg scores using subsequence DTW.  The most notable characteristic of our system is that it has no trainable weights at all---only a small set of hyperparameters that can be easily tuned on a small training set.  Our system generalizes very well from training to testing, and it achieves a test F measure score of $.869$.  We hope that this work serves as an entry point to exploring new ways to retrieve various forms of music using cell phone images as a query.

\section{Acknowledgments}
We would like to thank the Class of 1989 Summer Experiential Learning Fund, the Vandiver Summer Experiential Learning Fund, and the Norman F. Sprague III, M.D. Experiential Learning Fund established by the Jean Perkins Foundation for their generous support.  

\bibliography{SheetMidiRetrieval}

\end{document}